\documentclass[aps,letterpaper,amssymb,twocolumn,prl]{revtex4}
\usepackage{amssymb}
\usepackage[usenames,dvips]{color}
\usepackage{graphicx}
\usepackage{amsmath}
\usepackage{times}
\usepackage{subfigure}
\usepackage{times}
\begin{document}
\title{Particle-Hole Asymmetry and Brightening of 
Solitons in a Strongly Repulsive BEC}
\author{ Radha Balakrishnan$^{1}$, Indubala I Satija$^{2,3}$, and Charles W. Clark$^{3}$}
\affiliation{$^{1}$ The Institute of Mathematical Sciences, Chennai 600113, India}
 \affiliation{$^{2}$ Department of Physics, George Mason University, 
 Fairfax, VA 22030}
\affiliation{ $^{3}$ National Institute of Standards and
Technology, Gaithersburg, MD 20899}
\date{\today}
\begin{abstract}

We study solitary wave propagation in the condensate of a system of hard-core
bosons with nearest-neighbor interactions. For this strongly repulsive system, the evolution equation for the condensate order parameter of the system,
obtained using spin coherent state  averages is different from the usual Gross-Pitaevskii
equation (GPE).
The system is found to support two kinds  of solitons when there is a particle-hole imbalance:
a  dark soliton that dies out
as the velocity approaches the sound velocity, and a new 
type  of soliton which 
 brightens and persists all the way up to the
sound velocity,  
transforming into a periodic wave train at supersonic speed.
Analogous to the GPE soliton, the energy-momentum dispersion for both solitons is characterized by Lieb II modes.

\end{abstract}
\maketitle
Strongly correlated quantum systems pose some of the most difficult challenges at the forefront of fundamental physics.
Recent  theoretical and  experimental work in the field of Bose-Einstein condensates (BEC) of atomic gases  
 center around unveiling new phenomena that could lead to the
   understanding of  various complexities of these systems.
The possibility of tuning  inter-atomic interactions via Feshbach resonances
allows one  to  manipulate nature, and study
realistic and tractable quantum  many-body models.
Among the various models, a system of impenetrable bosons known as the  
hard-core boson (HCB) gas, is a paradigm. It was 
analyzed exactly in one dimension by Girardeau\cite{G}.  
 It has also been used
to explore quantum phase transitions\cite{Mott} and transport characteristics of bosonic and 
spin-polarized fermionic atoms\cite{GW}.

In this paper, we study 
solitary wave propagation
in a HCB system with nearest-neighbor (nn) interactions\cite{Mats} to 
obtain  deeper insight into beyond-GPE dynamics in 
quantum many-body systems.
A soliton or solitary  wave is a  localized nonlinear excitation  that travels
with a constant speed, retaining its shape.
Non-dispersive solitonic energy transport 
has been observed in a
BEC\cite{GPEdark}. These are the
well known dark solitons
that travel with speeds less than that of sound. 
Such particle-like transport is an active field 
with particular emphasis on unveiling many-body characteristics
that cannot be described with the approximate description 
provided by the GPE. The existence of dark solitons have also been shown in
a one-dimensional HCB gas using Fermi-Bose mapping \cite{GW}, 
and in a generalized mean-field theory where
the cubic nonlinearity of the GPE was replaced by a quintic term\cite{Kolo}.
Solitons in similar quintic models were further investigated with a periodic potential\cite{Alf},
and also in the presence of a dipole-dipole interaction.\cite{Baiz}
Additionally, various numerical investigations\cite{Carr}
have analyzed the quantum dynamics of dark solitons
to study the effects of quantum fluctuations 
and quantum depletion in  a Bose-Hubbard model.

Our formulation, based on
the equation for the BEC order parameter
obtained using spin coherent state averages\cite{radha1} 
differs considerably from earlier studies.
In addition to not being restricted to one-dimension, the evolution equation for the order parameter 
contains all powers of the condensate density. In this non-GPE type equation, which incorporates  quantum fluctuations
and depletion,
both particles and holes emerge as 
equal partners in the transport. 
In this regard, our methodology bears some parallel 
with recent work\cite{Mitra}
on  weakly-interacting atoms where a 
non-GPE description emerges due to quantum fluctuations.

In contrast to the weakly repulsive condensate described by 
the GPE which supports only a dark soliton\cite{Pitabook}, 
we show that  the HCB  condensate 
supports two distinct types of solitons: a  
dark  soliton whose amplitude vanishes as
its propagation speed approaches 
the Bogoliubov  speed of sound $v_s$
(like the GPE soliton),  and
a new species  of soliton that
exhibits  brightening and {\it persists} all the way up to $v_s$. 
Hence we call this a {\it persistent} soliton. 
For  propagation speeds above $v_s$, 
it  evolves into a periodic wave train. 
The existence of this novel type of soliton 
that brightens the condensate profile
is tied to the particle-hole population imbalance, a key parameter
for the cross-over to  a non-GPE behavior.

It is instructive to start our analysis with the
extended lattice Bose-Hubbard model in 
$d$ dimensions, 
\begin{equation}
H=-\sum_{j,a}[t \,b_j^{\dagger} b_{j+a}+ V n_j n_{j+a}]\\+\sum_j U n_{j}  
(n_{j}-1) 
-(\mu-2t)n_{j}
\label {BH}
\end{equation}
Here,  $b_j^{\dagger}$ and $b_j$ are the creation and  
annihilation operators 
for a  boson at the lattice site $j$,  $n_{j}$ is the number operator, $a$ 
labels nearest-neighbor (nn) sites , 
$t$ is the nn hopping parameter,   $U$ is the on-site repulsion strength, 
and $\mu$ is the chemical potential. 
To soften the effect of strong onsite repulsion, we add
 an attractive nn interaction ($V > 0$), (although our results will be valid for
 repulsive $V$). Such a term 
may mimic 
certain aspects of the long range dipole-dipole interaction, that 
are subject of various recent investigations.\cite{Baiz} 
The on-site term $2tn_j$ is added, so that the terms with $t$ 
reduce to the kinetic energy 
in the continuum version of the many-body bosonic Hamiltonian.

The HCB limit  ($U \rightarrow \infty$)  of the strongly  repulsive 
Bose-Hubbard model 
corresponds to 
the constraint that two bosons  cannot occupy the same site. 
This can be incorporated in the formulation by using 
field operators that anticommute at same site but commute at
different sites, thus satisfying
the same algebra
as that of a spin-$\frac{1}{2}$ system. 
The system can be mapped to
the following quantum XXZ Hamiltonian in a magnetic field\cite{Sach}  
by identifying $b_j$ with the spin flip operator 
$S_{j}^{+}$, along with 
$n_{j} = \frac{1}{2} - S_{j}^{z}$,  
\begin{eqnarray}
H_S=-\sum_{j, a}[tS_j^+S_{j+a}^-+V S_j^z S_{j+a}^z]-
\sum_j({\textstyle}U_{e}-\mu)S_j^z.
\label{spinH}
\end{eqnarray}
Here $U_{e}=(t-V)d$,  where $d$ is the spatial dimensionality.

The dynamics of the HCB system (\ref{spinH}) is described by the Heisenberg  equation of motion,
\begin{equation}
i \hbar \dot{S}_j^+ 
=({\textstyle}U_{e}-\mu) 
S_j^+ - t S_j^z\sum_{a}S_{j+a}^+ + V S_j^+ \sum_{a} S_{j+a}^z .
\label{Heis}
 \end{equation}
 This operator equation can be  
 transformed into an equation of motion for the  condensate order parameter $\eta_j=\langle S_j^+\rangle$
using spin-coherent state averages\cite{radha1}, a natural choice for 
describing the inherent coherence in the condensed phase of HCB.
Parameterizing the local order parameter as
$\eta_j= \frac{1}{2} \sin \theta_j \,e^{i\phi_j}$,
we obtain the condensate number density $\rho^c_j=|\eta_j|^2=
\frac{1}{4} \sin^2 \theta_j$ and
the  particle number density
$\rho_j=\langle n_j\rangle=\langle S_j^- S_j^+\rangle
=\sin^2 (\frac{1}{2}\theta_j)$.  Hence, the condensate and  particle
number densities are related by
\begin{equation}
\rho^c_j=\rho_j (1-\rho_j).
\label {den}
\end{equation}
Therefore, in contrast to the GPE,  we now have $\rho_j 
\not\equiv 
\rho^c_j$.
Further, the  resulting  formulation  encodes   fluctuations 
and depletion, as seen from the relations 
$\langle S_j^-S_j^+\rangle -\langle S_j^- \rangle 
\langle S_j^+\rangle 
 = \rho_j^2$ and
$\langle n_j^2\rangle - \langle n_j\rangle ^2=\rho_j^c$.

The Hamiltonian in Eq. (\ref{BH}) is invariant, 
up to a change in the chemical potential,
 under a particle-hole transformation,
 where the hole operators are the hermitian conjugates of the boson operators,
and the hole density is
$\rho^h_j = 1-\rho_j$. Thus  the condensate 
density  $\rho_j^{c}$ is the product of the particle density and the
hole density (from Eq. (\ref{den})), with
particles and holes playing equal roles in determining the condensate fraction.
As we now show, the particle-hole imbalance variable $\delta_j=
 (1 - 2 \rho_j) = \rho^h_j-\rho_j$ plays a key role in the dynamical evolution of the system.
 
The equations of motion for $S_j^+$ and $S_j^z$ lead 
to the following coupled equations for $\eta_j$ and $\delta_j$:
\begin{eqnarray}
i \hbar\,\dot{\eta}_j&=&({\textstyle}U_{e}-\mu)\eta_j 
- {\textstyle}t \,\delta_j
\sum_{a}\eta_{j+a}
+{\textstyle\frac{1}{2}}V\eta_j \sum_{a}\delta_{j+a}, \nonumber \\
i\hbar \dot{\delta}_j&=&2t \sum_{a} 
\big(\eta_j\,
\eta^*_{j+a}-\eta^*_j\,\eta_{j+a}\big).
\label{discreteeqs}
\end{eqnarray}
We now consider the continuum approximation
of the discrete equations (\ref{discreteeqs}), 
useful in the limit when the order parameter is a  smoothly varying
function with  a length scale greater than the lattice spacing $a$.
In the limit where the number of particles $N$ and the number of lattice sites $L$ 
tend to infinity,   with the 
filling factor $N/L$ fixed, the system is described by the
condensate order parameter $\eta(\bf{r})  
= \frac{1}{2} \sin \theta(\bf{r}) \,e^{i\phi(\bf{r})}$,
which is coupled to the 
particle-hole imbalance  variable
$\delta(\bf{r})=\cos \theta(\bf{r})$:

\begin{figure}[htbp]
\includegraphics[width =1.0\linewidth,height=1\linewidth]{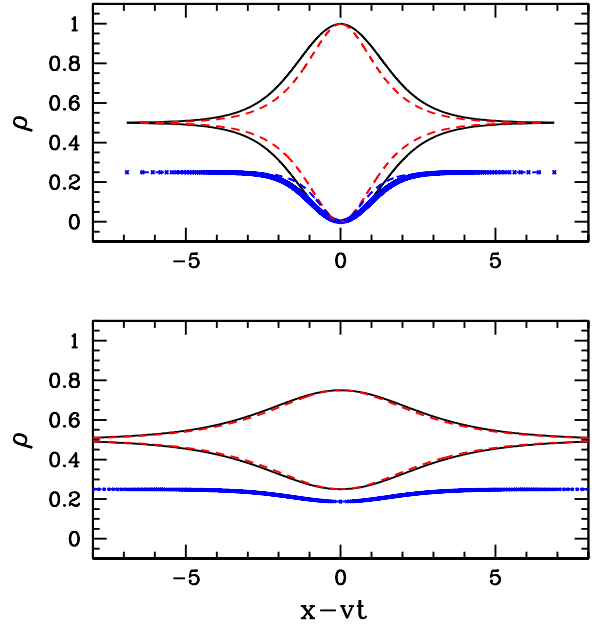}
\leavevmode \caption{(color online) Comparison of numerical (solid line:  black) and analytical (dashed line:  red)  
solitary wave solutions,
for $\rho_0 = 0.5$. The upper and lower panels  correspond to
$\bar{v} = 0, \gamma = 1$ and $\bar{v} = 0.87, \gamma = 0.5$, respectively.
The plots with crosses (blue) show the corresponding condensate density $\rho^{c}$.}
\label{fig1}
\end{figure}

\begin{figure}[htbp]
\includegraphics[width =1.0\linewidth,height=1\linewidth]{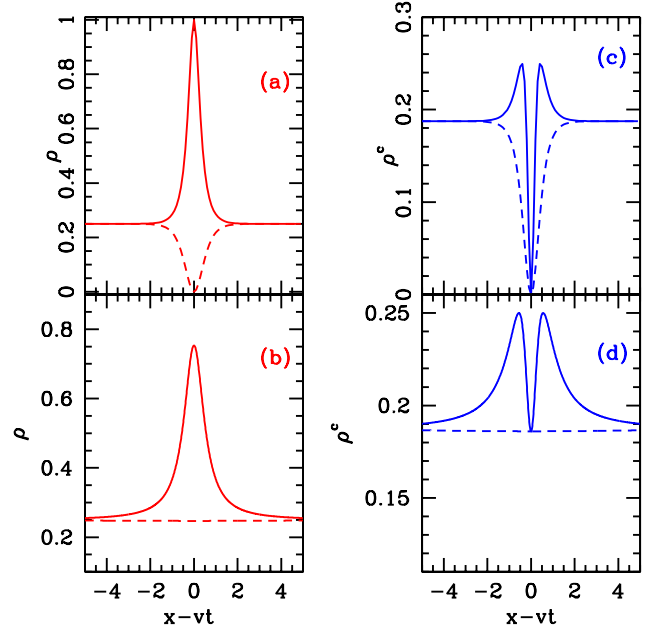}
\leavevmode \caption{(color online) Solitary wave for $\rho_0 = 0.25$ for
total particle density (a,b) and condensate density (c,d), for
 $\bar{v}= 0$ (a,c) and $\bar{v}=1$ (b,d).
Solid (dotted) plots correspond to $f^+$ ($f^-$) solitons. }
\label{fig2}
\end{figure}

\begin{equation}
i \hbar \dot{\eta} 
= -\frac{\hbar^2}{2m} \delta \,\,\,\nabla^2 \eta 
+\frac{V_e}{2} \eta \,\,\,\nabla^2 \delta
+ \frac{1}{2} U_e(1 - \delta) \eta-\mu \eta, \\ 
\label{orderpar}
\end{equation}
\begin{equation}
\dot{\delta}=\frac{\hbar}{2m} \nabla\cdot [(1-\delta^2)\nabla \phi],
\label{cont}
\end{equation}
where $ta^2=\hbar^2/(2m)$ and $V_e=Va^2$.

In the small-$\eta$ limit where we neglect terms involving $|\eta|^{2n}\eta$ ($n > 1$), and retain only linear terms involving
the derivatives of $\eta$,  both the order 
parameter equation (\ref{orderpar}) and the continuity equation
(\ref{cont}) reduce to the  GPE form,
with an  effective mean field interaction  equal to $U_{e}$.
In the above equations, if we set $(1 - \delta) = 2\rho $ and use 
Eq. (\ref{den}), we can show that
for $\rho({\bf r}) < \frac{1}{2}$, the GPE limit is obtained when
 $\rho^c \approx \rho$, i.e., the order parameter corresponds to
 the condensate of particles,
 while for $\rho({\bf r}) > \frac{1}{2}$, the GPE  is satisfied by 
 the order parameter for the condensate of holes,
namely  $\eta^*$,  with $\rho^c \approx \rho^h$.

The Bogoliubov spectrum associated with 
the {\it small amplitude}  modes  of the HCB system 
is similar to the GPE\cite{radha1} case and determines the
Bogoliubov sound velocity $v_s = (U_{e}\rho^c/m)^{1/2}$.
However, as we discuss below, for the propagation 
of {\it nonlinear} localized modes, true GPE-type 
transport emerges only near half-filling and
the crossover to non-GPE dynamics is controlled 
by the particle-hole imbalance.
\begin{figure}[htbp]
\includegraphics[width =.8\linewidth,height=.5\linewidth]{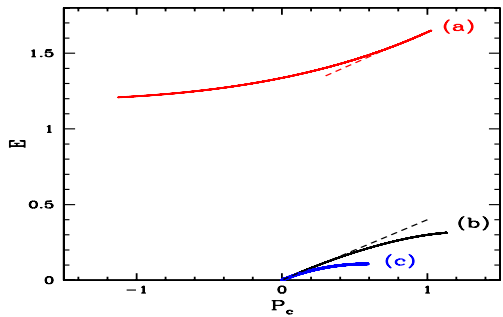}
\leavevmode \caption{(color online) Energy vs canonical momentum for 
$\rho_{0} = 0.25$ for nn interaction $V=0$.
Branches  (a) and  (b)  correspond to the persistent soliton and the dark soliton, respectively.
Curve (c) gives the corresponding dispersion for the GPE soliton with $\rho=\rho^c_0$.
The dashed tangents on the two branches demonstrate  
linear dispersion, with  identical slopes.}
\label{fig3}
\end{figure}

We now investigate unidirectional soliton propagation in the 
HCB system, described by 
Eqs. (\ref{orderpar}) and (\ref{cont}). 
To this end, we 
look for traveling wave solutions for $\rho$  with velocity $v$, of the form
$\rho(z)=\rho_0+f(z)$, where $z=(x-vt)/a$ and $\rho_0$ denotes  a uniform background density. 
The corresponding background particle-hole imbalance 
is $\delta_0 =  (1-2\rho_0)$. 
The function $f(z)$ is then found to satisfy the nonlinear equation
\begin{equation}
4\left(\frac{df}{d\bar{z}}\right)^2\left[1-\zeta^2 \frac{f(f-\delta_0)}{\rho_0^c}
\right]=
f^2\left[\gamma^2-\frac{f(f-\delta_0)}{\rho_0^c}\right],
\label{feqn}
\end{equation}
where $\gamma^2 = 1-\bar{v}^2$ with $\bar{v}=v/v_s$. Further, 
$\bar{z}=\zeta z$ and $\zeta = \Lambda/\sqrt{1-\Lambda^2}$, 
where the microscopic parameter $\Lambda= v_s\,ma/\hbar$ 
is the Bogoliubov speed of sound $v_s$
measured in units of the zero-point velocity $\hbar/ma$. 

Equation  (\ref{feqn}) is satisfied by an 
elliptic integral. In the special case where the 
bracketed term on the left-hand side
 is approximated by  unity, there are two
soliton solutions $f^{\pm}$ of the form\cite{radha}
\begin{equation}
f^{\pm} = \frac{2\gamma^2\rho^c_0}{\pm
\sqrt{\delta_0^2+4\gamma^2 \rho^c_0} \,\,\cosh\,(z/\Gamma) - \delta_0}.
\label{ffunction}
\end{equation}
Here,  $\Gamma=2\zeta \gamma$ is the width of the soliton.
These two special solutions are emblematic 
of the general solution to  Eq. (\ref{feqn}).
We have numerically integrated Eq. (\ref{feqn}) to 
find $f(z)$, and compared it with the analytical solution (\ref{ffunction}).
We find close agreement between the numerical and 
analytical solutions for all values of $v$ and $\rho_0$. 
This  happens  because the neglected terms become
irrelevant for small $f$ and also when $\frac{df}{dz}=0$, the condition which determines almost all the contributions to the localized modes.
Fig.~1 with $\rho_0=1/2$
shows the case when the agreement is at its worst. Away from half-filling,  
the numerical and analytic solutions are very close.

The solutions (\ref{ffunction}) are valid provided $U_{e}>0$ (or $t > V$), 
because the functions $f^{\pm}$ are unbounded for negative $U_{e}$.
We point out that the interaction $V$ 
simply scales the width of the soliton, without altering its profile.
This leads to an interesting possibility that the existence of the soliton and its overall profile 
may remain unaffected by certain long range interactions.\cite{Baiz}

The distinction between the twin solitons, which encode particle-hole duality as
$f^+(z, \rho_0) = f^-(z, \rho_0^h)$, can be elucidated by computing the momentum
$P^{\pm}= m \int J^{\pm} dx$. 
Here $J= (\hbar/m)\rho_c\nabla \phi$ 
(see Eq. (\ref{cont})) is the condensate current, which
is obtained using the equation\cite{radha}  $d\phi/dz = 
vma\,(\rho^{\pm}~-~ \rho_{0} )/\hbar \rho_c$. We get 
\begin{equation}
P^{\pm}= \pm (mv) \frac{\sqrt{\rho_c^0}}{2\zeta} \tan^{-1}\sqrt{\frac{(\delta_0^2+2\rho_c^0\gamma^2)^{1/2}\pm
\delta_0}{(\delta_0^2 + 2\rho_c^0\gamma^2)^{1/2}\mp\delta_0}}\,.
\label{mom}
\end{equation}
For small $v$, 
$P^{+}$ corresponds to the momentum of a 
particle of positive mass while $P^-$ corresponds to the momentum of a hole,
as the effective mass is negative. This justifies associating $f^+$ 
 [respectively, $f^-$] 
as the solitary wave with particles [holes]. 
Further,  $(P^+-P^-)/mv=\pi \zeta/\sqrt{\rho^c_0}$, which is suggestive of a conservation principle.

Particle-hole imbalance appears to be a key factor in determining the characteristics of the solitary waves.
As seen in  Fig. ~1, when the number of particles is equal to 
the number of holes,
the two solutions for the particle density $\rho(z)$ are dark and anti-dark\cite{anti} solitons ( bright solitons on a pedestal)
 which are mirror images of each other. 
In the corresponding
condensate density $\rho^c(z)$, the solutions are indistinguishable, resulting in a single GPE-like soliton, 
that flattens out at the sound velocity. In 
fact, within the approximations discussed above, 
Eq. (\ref{feqn}) expressed in terms of the condensate density
$\rho_c(z)$ 
is identical to a GPE with healing length $\xi$ 
renormalized to $\frac{1}{2}\xi \sqrt{1-\Lambda^2}$.
In other words, when the  background consists of equal
numbers of particles and holes, the soliton solution for the condensate density of the HCB is like  
the usual soliton of the GPE, but with a renormalized healing length.

When the number of particles differs from the number of holes, 
the condensate soliton corresponding to the solution 
$f^{-}$ is dark and flattens out at the sound velocity.
However, the condensate soliton corresponding to 
the solution $f^{+}$ brightens the condensate profile,
as shown in Fig. 2.
As $v$ increases, the spatial extent of the disturbance above the background increases, and the solitary wave $\rho^c(z)$  becomes completely 
bright at the speed of sound. More important, this soliton does not 
flatten out but  persists even at the sound velocity.
The survival or persistence of the soliton of this sector 
is evident from the limiting functional form\cite{sky} 
of $f^+$ as $\bar{v} \rightarrow 1$. 
We find  
$\rho^+ \rightarrow 
\rho_0+ \delta_0/(1+Cz^2)$,
where   $C=4\zeta^2 \delta_0/\rho^c_0$ and hence the width of the soliton becomes independent of its speed of propagation.
It should be noted that as $\rho_0 \rightarrow 0$, the solutions approach a definite limit with 
$C$ equals to $8U_e$ in units of zero-point energy $(\hbar/a)^2/m$. 
In other words, the anti-dark solitons become bright solitons,
reminiscent of the bright solitons discussed in systems with attractive dipole-dipole interaction.\cite{Baiz}

Another distinctive feature of the persistent soliton is that it is transformed into
a periodic  wave train  when its velocity exceeds the sound velocity.
Indeed, inspection of Eq. (\ref{ffunction}) shows that  at supersonic speeds, 
when $\gamma$ becomes imaginary, 
the  soliton is transformed to
a spatially periodic wave that exists
for velocities $ v_s < v  <  v_s/(1-\delta_0^2)$.

Finally, we compute the energy-momentum 
relationship of the solitons by integrating the equation 
$dE/dP_c = v$ to obtain the canonical 
momentum $P_c$, where $E$
is the energy\cite{Pitabook}. 
The three plots in Fig. ~3 show the dispersion relations for
the $f^+$  and $f^-$  solitons, and the corresponding GPE soliton\cite{BC}.
Each plot  shows linear dispersion at one end of the spectrum (at the momentum
corresponding to $v=v_s$), 
and saturation at the other end (at the momentum corresponding to $v=0$). 
The dark soliton has a {\it linear}
dispersion near $ E =0, P_c =0$), with the slope given by  $v_s$, and it saturates to zero slope 
near the maximum value of the momentum. (The GPE  soliton, with $\rho = \rho_{0}^{c}$,
has a similar behavior). The persistent  soliton
exhibits {\it linear} dispersion with the {\it same} slope $v_s$ near the maximum value of the momentum.
Comparing this with the exact bosonic low-lying excitation spectrum for the HCB gas discussed by Lieb \cite{Lieb}, 
we note that both soliton branches with  bounded momentum intervals, mimic the main characteristics of his type II excitation spectrum, being linear at one end and saturating at the other. 
However, it should be noted that in the present context,
it is the lower branch that can be designated as a type II
`hole' state.  The higher energy upper branch associated 
with the persistent soliton is in fact a type II `particle' state.
This is different from Lieb's classification, where `particle' states 
were always type I with unbounded momentum.

In summary, we have explored solitary wave 
propagation in a degenerate HCB gas, and have found 
novel features that are not present in
the conventional
Gross-Pitaevskii equation, as well as some others that are. 
In a lattice system, this cross-over occurs at  half-filling, which corresponds to
equal numbers of particles and holes in the HCB  system. 
At exact half-filling, we obtain GPE-like  dark solitary waves. 
Away from half-filling, we find both dark and 
anti-dark solitary waves including 
a persistent variety 
which propagates with a non-vanishing  amplitude 
right up to  
the speed of the sound. 
The forms of these solutions at $t =0$ represent 
two initial types of 
disturbance profiles that could evolve into  these solitons in realistic quasi-one-dimensional systems. These aspects could be further 
explored in numerical simulation in first-principle quantum many-body calculations and may find experimental realization in 
a highly anisotropic cigar shaped trap.
We hope that our results will also provide further stimulus to the study of solitons in quantum many-body systems.


\begin{thebibliography}{99}
\bibitem{G} M. Girardeau, J. Math. Phys. {\bf 1}, 516 (1960).

\bibitem{Mott} F. Herbet {\it et al.}, Phys. Rev. B {\bf 65}, 014513 (2002).

\bibitem{GW} M. D. Girardeau and E. M. Wright, Phys. Rev. Lett.  {\bf 84}, 5691  (2000);
Phys. Rev. A {\bf 77}, 043612 (2008).

\bibitem{Mats} T.  Matsubara and H. Matsuda, Prog. Theor. Phys. {\bf 16},  569 (1956).

\bibitem{GPEdark} S. Burger {\em et al.}, Phys. Rev. Lett. {\bf 83}, 5198 (1999); 
J. Denschlag {\em et al.}, Science {\bf 287}, 97 (2000)

\bibitem{Kolo} E. Kolomeisky, T. J. Newman, J. Straley, 
and X. Qi,  Phys. Rev. Lett. {\bf 85}, 1146 (2000).

\bibitem{Alf} G. L. Alfimov {\em et al}., Phys Rev. A, {\bf 75}, 023624 (2007).

\bibitem{Baiz} B. B. Baizakov {\em et al}. J. Phys. B: At. Mol. Opt. Phys., {\bf 42} (2009) 175302;
G. Gligoric {\em et al}, Phys Rev A, {\bf 78} 063615, (2008).

\bibitem{Carr} K. V. Krutitsky {\em et al}., arXiv: 0907.0625 (2009); V. Mishmash et al., arXiv: 0906.4949 (2009).

\bibitem{Sach} See for example, S.Sachdev {\it Quantum Phase Transitions}, (Cambridge University Press, Cambridge, 1999).

\bibitem{radha1} R. Balakrishnan, R. Sridhar, and R. Vasudevan, 
Phys. Rev. B {\bf 39} 174, (1989).

\bibitem{Mitra} K. Mitra, C. J. Williams, and C.A.R. Sa de Melo,  Phys. Rev. A, {\bf 77}, 033607 (2008).

\bibitem{Pitabook} L. Pitaevskii and S. Stringari, {\it Bose-Einstein Condensation} (Oxford University Press, Oxford,  2003).


\bibitem{radha} R. Balakrishnan,  Phys. Rev. B  {\bf 42}, 6153 (1990).

\bibitem{anti} Y. S. Kivshar and V. V. Afanasjev, Phys. Rev. A {\bf 44}, R1446 (1991).

\bibitem{sky} This solution is
reminiscent of the skyrmion in 2D:
with $\rho(z=0)=
1-\rho(z\rightarrow \pm\infty)$, these solitons correspond 
to configurations with up spins at $r \rightarrow \infty$ and 
down spins at $r \rightarrow 0$.

\bibitem{BC} The integration is done using the boundary condition $P_c^-=0$
when $v=v_s$ for the $f^{-}$ soliton. For the $f^{+}$ soliton, we use 
 the condition $P_c^+(v=0)=-P_c^-(v=0)$,  because 
 the twin solitons have equal and opposite phase jumps at $v=0$. 

\bibitem{Lieb} E. H. Lieb, Phys Rev, {\bf 130} (1963), 1616.

\end{thebibliography}
\end{document}